\documentclass{emulateapj}

\newcommand\LCDM{$\Lambda$CDM\ }

\newcommand\msun{\rm M_\odot}
\newcommand{\cc}{{\rm cm}^{-3}}
\newcommand{\nH}{n_{\rm H}}
\newcommand{\nHp}{n_{{\rm H}^+}}
\newcommand{\nHm}{n_{{\rm H}^-}}
\newcommand{\nHH}{n_{{\rm H}_2}}
\newcommand{\nHHp}{n_{{\rm H}_2^+}}
\newcommand{\nD}{n_{\rm D}}
\newcommand{\nDp}{n_{{\rm D}^+}}
\newcommand{\nHD}{n_{{\rm HD}}}
\newcommand{\enzo}{Enzo}

\shorttitle{HD Chemistry in the First Stars}
\shortauthors{McGreer and Bryan}

\begin{document}

\title{The Impact of HD Cooling on the Formation of the First Stars}
\author{Ian D. McGreer and Greg L. Bryan}
\affil{Department of Astronomy, Columbia University, Pupin Physics Laboratories, New York, NY 10027}

\begin{abstract}

We use numerical simulations to investigate the importance of HD formation 
and cooling on the first generation of metal-free stars in a \LCDM cosmology.  
We have implemented and tested non-equilibrium HD chemistry in an adaptive
mesh refinement simulation code and applied it to two situations.   (1) It is first
applied to the formation of $10^5 - 10^6$ $\msun$ halos which form in the
absence of any ionizing source (``unperturbed'' halos).  We show, in agreement 
with previous work, that HD cooling is of only marginal importance for most 
halos; however, we find that for the lowest mass halos, with masses a few 
times $10^5 \msun$, HD cooling can equal or surpass the H$_2$ cooling rate.
This leads to a population of stars formed in halos with effective HD cooling 
that are less massive by a factor of $\sim 6$ compared to halos dominated 
by H$_2$ cooling.  (2) In the second part of the paper, we ionize the halos 
in order to explore the impact of HD cooling in the presence of an ample 
population of free electrons.  This leads to cooler temperatures (due to the 
electron-catalyzed production of H$_2$) implying somewhat lower resulting 
proto-stellar mass.  Adding HD chemistry changes this by lowering the 
temperature further, to the level of the CMB.  We find that HD cooling 
dominates over H$_2$ cooling in the density range $10^2$ cm$^{-3}$ 
to $10^6$ cm$^{-3}$, but above this density, the temperature rises  
and H$_2$ cooling dominates again.  Because of this, the accretion rate on to 
the protostar is almost the same as in the H$_2$ case (at least for accreted 
masses below 50-100 $\msun$), therefore we argue that HD cooling in ionized 
halos will probably not result in a population of significantly lower mass stars.

\end{abstract}


\section{Introduction}

The conditions surrounding the formation of the first stars out of 
primordial gas has been the focus of much work over the last few years, 
and has recently been summarized by \citet{brommlarson04} and 
\citet{ciardiferrara05}.  The picture that emerges is that of the 
first stars forming in halos with masses around $10^6$ $\msun$ 
due to cooling from molecular hydrogen 
\citep[e.g.,][]{couchman86, haiman96, tegmark97,  abn02, yoshida03}.  
The actual masses of the first stars are less well determined, but in 
the last five years a number of groups have employed high-resolution 
hydrodynamic simulations, using both adaptive Eulerian codes 
\citep[e.g.,][]{abn02, oshea07} and smoothed particle hydrodynamics 
(SPH) Lagrangian codes \citep[e.g.,][]{bcl02, yoshida03,yoshida06,gao07} 
to answer this question.  This work has generally resulted in the formation 
of a single, relatively high mass (of order $100 \msun$) star forming in each 
halo. The resulting ionization front from the first star ionizes the original cloud 
before any other stars can form \citep{whalen04, kitayama04, abel07}.

Much of the early work focused on single objects that formed within 
the most massive halo in the simulation volume.  More recently, some 
groups have explored the parameter space by conducting multiple 
simulations with different initial conditions and box sizes \citep{oshea07}
and with varying cosmological parameters \citep{gao07}.  In general, 
a fairly consistent picture results: massive stars form in overdense 
regions (halo masses $\sim 10^6 \msun$) less than a billion years 
after the big bang ($10 \la z \la 30$).

Most first star simulations have relied on a nine species chemical 
reaction network consisting of hydrogen, helium, molecular hydrogen, 
and their associated ions \citep[e.g.,][]{abel97}. This simple model is 
possible because big bang nucleosynthesis produced only lithium and
lighter elements. While only trace amounts of H$_2$ are present 
initially within the primordial gas, it forms rapidly in the high density 
gas within collapsing halos, and once the density exceeds 
$n\sim10^8~\mathrm{cm}^{-3}$ the three-body formation process 
drives the fraction of molecular hydrogen to near unity.

H$_2$ has been shown to be effective in cooling primordial 
gas within dense halos, lowering the gas temperature to $\sim 200$K.
However, the H$_2$ molecule itself is a relatively poor coolant since 
it lacks a permanent electric dipole moment.  On the other hand, the 
HD molecule has a permanent dipole, and cools more effectively 
at low temperatures ($\la 200$K). This has led several authors to 
suggest that HD will have an important effect on first star formation 
\citep{puy93, gp98, stancil98}.  

However, in low-mass primordial halos, the temperature never falls 
below about 200K, which is required for HD to dominate over H$_2$ 
cooling, and so the general consensus has been that HD is unimportant 
for primordial, low-mass halos.  For example, \citet{bcl02} included HD
 chemistry in 3D SPH simulations of first star formation and found that 
H$_2$ cooling dominated; \citet{yoshida06} came to the same conclusion. 
\citet{ripamonti07} recently conducted a suite of 1D simulations with full 
HD chemistry and found that for typical halos this is true, but that in low 
mass halos ($\la 3 \times 10^5 \msun$) HD is formed in sufficient quantities 
to be as important in cooling the gas as H$_2$. In one run, HD cooling lowered 
the temperature of the gas to $\sim70 K$, 3-4 times lower than when HD 
was neglected. \citet{ripamonti07} argued that the Jeans mass in such halos 
is lowered by a factor of 10, implying that significantly lower mass stars will 
form in these regions.  This is an intriguing suggestion and one of the 
motivations of this paper is to test this idea using full three-dimensional 
simulations.

While there is general (but not universal) agreement that HD is likely to 
be mostly unimportant in low-mass halos that have never been ionized, 
a number of authors have recently pointed out that HD will form in high 
abundance when the gas is ionized 
\citep[as originally pointed out by][]{sk87}.  For example, \citet{no05} 
used one-dimensional models to follow the evolution of HII regions inside 
massive halos.  They found that for high enough masses, the gas would 
re-collapse and form significant amounts of HD due to the presence of 
free electrons.  This cooling lowered the gas temperature to the CMB 
temperature and so the authors suggested that low mass stars (of 
order 1 $\msun$) could result.  \citet{machida05} also looked at the 
impact of HD cooling in the swept up shells of primordial supernovae 
\citep[see also][]{salvaterra04}.

More generally, \cite{jb06} used one zone models to show that HD 
cooling would be important whenever primordial gas was significantly 
ionized.  This included halos forming behind strong shocks, in relic HII 
regions and in high-mass halos.  They showed that the cooling was 
sufficient to quickly reach the CMB temperature and argued that this 
would give rise to a Population II.5\footnote{Stars which formed in
primordial, metal-free gas that has been perturbed by a previous
generation of stars have often been referred to as Population II.5; i.e.,
an intermediate generation between Population III and Population II.
A recent proposal \citep{firststars3} is to refer to the ``first generation'' 
stars formed in primordial gas as Population III.1, and the 
``second generation'' stars formed in gas of primordial composition
but affected by feedback effects from the first generation as Population III.2, 
a nomenclature we will adopt in this paper.} 
\citep{mackey03} with masses 10 times lower than traditional 
Population III stars.  \citet{gb06} went  on to suggest that these 
stars generated the majority of the photons that ionized the universe.

In this paper, we use three-dimensional simulations to explore the
impact of HD cooling in two related areas of primordial star formation.
The first is to investigate more carefully the role of HD cooling in 
``unperturbed" low-mass primordial star forming regions, paying 
particularly close attention to the suggestion of \citet{ripamonti07} 
that HD may play a key role in the lowest mass halos.

In the second part of this paper, we explore the impact that free 
electrons have on HD cooling in primordial halos.  As noted above, 
ionization can occur for a variety of reasons, including relic HII regions 
(the kpc-sized regions ionized by nearby primordial stars) and shocks 
due to mergers with higher mass halos.  

We take a deliberately simple approach to the generation 
of free electrons and we simply ionize and heat the gas instantaneously 
at a certain point in the evolution of our halos.  This is an approximation 
to more sophisticated models which follow the radiative transfer of the 
ionizing radiation and its heating explicitly.  Our method is very similar 
to \citet{oshea05}.  As discussed more fully in \citet{mbh06}, this 
approach has two distinct differences compared to a more realistic 
treatment.  First, the optically thin radiative cooling approximation means 
that the dense gas in the core may be improperly ionized (although if the 
source is in the center of halo itself, this is less of a concern).  Second, the 
instantaneous nature of the heating prevents the gas from being expelled from 
the halo. The advantage of our approach is that we are able to compare the 
effect of HD cooling with and without ionization in similar halos, and so we can 
see the key effects simply and clearly.  Also, it is computationally faster 
and so we can study more cases.  Finally, as we will argue later, many of 
our key results should still apply when carried out with more realistic 
treatment of the ionization (and indeed, do match other simulations 
where we can compare).

In the next section, we describe the simulation code and our 
treatment and tests of HD chemistry.  In \S\ref{sec:results1}, 
we first examine the impact of HD on primordial halos without 
ionization, and then in \S\ref{sec:results2} we analyze the 
role of HD cooling in halos with ionization. Section~\ref{sec:discussion}
contains a discussion of the general effects of HD cooling in primordial 
star formation, including predictions for the final masses of stars based
on the end state of the simulations.  Finally, a brief summary of the
results is presented in \S\ref{sec:conclusions}.

\section{Methodology}

To follow the formation and evolution of the primordial clouds from 
cosmological initial conditions, we use the \enzo\ adaptive mesh 
refinement (AMR) method.  It is a well-tested cosmological hydrodynamics 
scheme that has previously been used to study a number of astrophysical 
systems including clusters of galaxies \citep[e.g.,][]{Bryan2001, Younger2007}, 
the ISM \citep{Slyz2005}, the intergalactic medium \citep{Richter2006}, and 
other applications.  In particular, it has been widely used to study the formation 
of the first stars at high redshift 
\citep{abn00,abn02,Machacek2001,Machacek2003,abel07,oshea07}.

The code is detailed more fully in \citet{bryan97}, \citet{Bryan1999}, 
and \citet{OShea2004}, and so we only summarize it here.  The code 
relies on a hybrid particle-mesh solver \citep{efstathiou85}, using an 
N-body representation for the collisionless dark matter, and an Eulerian 
grid for the baryonic component.  \enzo\ includes a second-order accurate 
piecewise parabolic method solver \citep{Colella1984,bryan95} for solving 
the hydrodynamic equations. The resolution of the grid is increased over the
course of a simulation using the block-structured AMR technique 
\citep[see][]{bergercolella89}.  On every level and at each timestep, the state 
of the grid is examined to determine if any regions have exceeded critical 
values in baryonic density, dark matter density, or Jeans length.  If so, then the 
grid is refined in that region by adding more grid points. In this way, the most 
interesting regions of a collapsing medium receive the highest levels of refinement, 
and dynamic range large enough to correctly model the small-scale physics of 
cloud collapse in a cosmologically significant volume is achieved.

As our method and application is very similar to \citet{abn02} and 
\citet{Machacek2003}, we describe only the changes in detail\footnote{The code 
used is nearly identical to the public version of \enzo\ described in \citet{OShea2004}
and available at http://lca.ucsd.edu/software/enzo/.}.
In the next two sections, we discuss first our enhanced chemical model 
and then the details of the simulation setup.

\subsection{Chemical Model}\label{sec:chemistry}

\begin{figure}[!t]
 \epsscale{1.2}
 \plotone{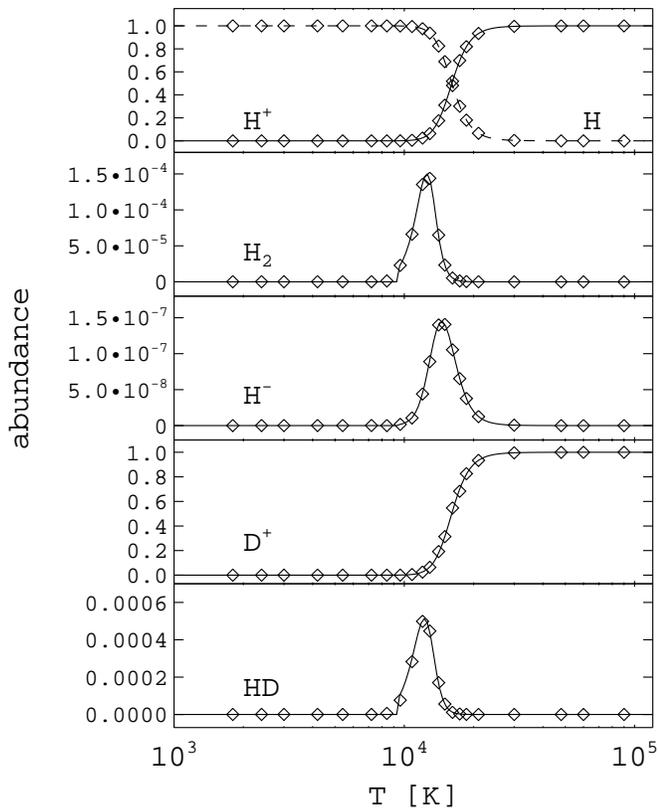}
 \caption{ 
  Equilibrium abundances of various species obtained from steady-state
  runs of \enzo. Lines show the analytic expectation from simple derivations
  based on the reaction rates (see Appendix). Diamond 
  symbols mark the values converged on by adiabatic runs of \enzo (with all
  radiative cooling turned off), showing that the chemistry solver reproduces 
  the expected values.
 \label{fig:equilibrium}
 }
\end{figure}

The first stars formed in a chemically simplified environment consisting of
lithium and lighter elements. Cooling from atomic hydrogen and helium is
inefficient in the low temperature ($\la 2000$K) gas within collapsing
halos at $z\sim20$, but should enough molecular hydrogen form
it can cool the gas to T $\sim 200$K. \citet{abel97} and \citet{anninos97} 
developed a chemical model for primordial gas following the evolution 
of the nine dominant species H, H$^{+}$, H$^{-}$, He, He$^{+}$, 
He$^{++}$, H$_2$, H$_2^+$, and e$^{-}$. \citet{abel97} 
demonstrated that the reaction rates for these species can be reduced 
to 19 collisional and 9 radiative processes and remain valid over a 
wide range of temperatures ($1< \mathrm{T} < 10^8$K) and densities 
($n \la 10^4 \mathrm{cm}^{-3}$). The rate equations are solved 
using an implicit scheme based on the backward differencing formula 
(BDF) technique described in \citet{anninos97}.  \citet{abn00} modeled 
the formation of a primordial molecular cloud using this chemical model, 
following the cloud fragmentation and collapse until a density of 
$10^5 \mathrm{cm}^{-3}$ and molecular hydrogen abundance of 
$10^{-4}$ was reached. In \citet{abn02} the formation of molecular 
hydrogen via the three-body process was added, which extended the range 
of valid densities to $n \la 10^{10} \mathrm{cm}^{-3}$ and allowed 
the collapse to continue until the formation of a fully molecular, 
$\sim 1 \msun$ protostar. No heating term for three-body H$_2$ formation
is included, an effect which can be important for the temperature
evolution of the gas at $n\ga10^{9}~\cc$.  In this paper we focus on
an intermediate regime at densities $\sim10^2-10^6~\cc$ where the
temperature is lowest. This regime is not sensitive to the high density 
core, and though we allow the simulations to continue to very high 
densities ($\sim10^{13}~\cc$) we do not consider the core in our analysis.

In this work, we introduce three additional species to the network: 
D, D$^{+}$, and HD. Reaction rates for the deuterium species are listed 
in Table~\ref{tab:rates}. A total of eight reactions are considered for the 
deuterium species. Five of the six reactions of the {\sl minimal model} of 
\citet{gp98} are included, using the updated rates from \citet{gp02}. 
We neglect photoionization of deuterium as we do not consider any 
external radiation fields other than the CMB, which does not contribute
to this rate at $z<100$. We include two additional deuterium reactions
from \citet{gp02} that were not part of the \citet{gp98} minimal model.  
Lastly, we include deuterium ionization by collisions with electrons, for 
which the rate is unknown and the equivalent hydrogen rate is adopted 
\citep[see, e.g.,][]{glover07a}.

The cosmological deuterium abundance is taken from \citet{burlestytler98} 
and is ${\rm D}/{\rm H} = 3.4\times10^{-5}$ (by number).  This is $\sim40\%$ 
larger than the value obtained by \citet{romano03} using WMAP first-year results, 
though it is within the range of other D/H measurements 
\citep[see][Table 1]{romano03}.  

We adopt the recently computed HD cooling function of \citet{lipovka}.
Their rate calculation includes all radiative and collisional transitions for
the $J \le 8$ rotational levels and the $v=0,1,2,3$ vibrational levels. The
rates are similar to previous formulations (e.g., \citealt{flower00}) 
except at high densities and temperatures ($T\ga1000$K) where the
collisional ro-vibrational transitions provide added cooling. From
\citet{lipovka}, we use their equation 5 for the low-density limit, and
in the LTE approximation we use the polynomial fit from their equation 4 with 
a density of $n_{\rm H} = 10^6 \mathrm{cm}^{-3}$ (at which point the 
cooling function has saturated). The high and low density limits are
combined as
\begin{equation}
 \Lambda(n_{\rm H},T) = \frac{\Lambda^{({\rm LTE})}}
                         {1+\frac{\Lambda^{({\rm LTE})}}
                                      {n_{\rm H}\Lambda^{(n \rightarrow 0)}}}.
\end{equation}
We prevent cooling below the CMB temperature by setting the HD
cooling rate to zero when the temperature in a simulation cell falls
below the CMB temperature at the current redshift.

\begin{figure}[t]
 \epsscale{1.2}
 \plotone{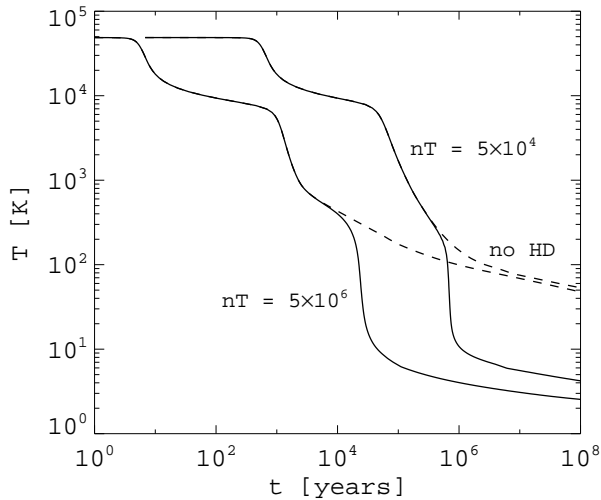}
 \caption{ 
  Thermal evolution of an ionized gas with an initial temperature of 50000K
  under isobaric conditions with and without HD cooling (solid and dashed
  lines, respectively). Once a temperature of $\la 300$K is reached, HD cooling 
  dominates and the gas cools to essentially to zero temperature, whereas gas 
  with H$_2$ cooling does not cool below $\sim 60$K. Note that this is not a 
  cosmological simulation and heating from the CMB was not considered. The 
  evolutionary phases of the gas are very similar to Figure 2 of \citet{yoshida07a}, 
  though the transitions occur on slightly different timescales due to the different 
  rates used.
 \label{fig:isobaric}
 }
\end{figure}

Following \citet{bcl02} and \citet{jb06}, we verified the chemical 
network in \enzo\ by comparing the equilibrium abundances of the species 
most important for cooling against values derived analytically. While \enzo~ is 
designed primarily for cosmological simulations, it is possible to set up idealized
problems and study the physics in detail. We created an adiabatic steady-state 
model consisting only of baryonic gas at an initial temperature and density 
uniformly distributed within a static (non-expanding) volume, and with radiative 
cooling disabled.  Using this setup, we ran simulations ranging in temperature 
from $10^3{\rm K} < T < 10^6{\rm K}$ and allowed each simulation to run 
until the abundances of all 12 species converged. The results are shown in 
Figure~\ref{fig:equilibrium}; the simulation data are in excellent agreement
with the values derived analytically in the appendix.

In addition to examining the equilibrium properties of an isothermal gas, we
considered the temperature evolution of hot gas that is cooling isobarically.
This simplified model approximates the conditions of photoionized gas
as may be expected for relic HII regions, SNe bubbles, or the shock ionized 
regions of colliding halos. Using the setup described above,
but with radiative cooling enabled and maintaining isobaric conditions by forcing 
$nT$ to be a constant at each timestep, we followed the temperature evolution of
the gas until a minimum temperature was reached. This test was motivated
by the work of \citet{yoshida07a}; comparison of our Figure~\ref{fig:isobaric}
to their Figure 2 shows that similar results are obtained by two different
numerical models. Figure~\ref{fig:isobaric} shows that HD cooling is important
in ionized gas, lowering the minimum temperature to $10$K (no radiative heating
from the CMB was considered in this run).

\subsection{Simulation Setup}\label{sec:setup}

The framework for our simulations is a single box 1 comoving Mpc 
on a side within a \LCDM universe with cosmological parameters 
consistent with \citet{wmap3}: $\Omega_\Lambda=0.76$, 
$\Omega_m=0.24$, $\Omega_b=0.041$, $H_0=73$~km/s/Mpc, 
and $\sigma_8=0.74$. The box is partitioned into a root grid of 
$128^3$ elements and initialized at $z=99$ with $128^3$ dark 
matter particles using an \citet{eisensteinhu99} power spectrum 
with a spectral index of $n=0.95$. The evolution of the dark matter 
particles is followed to $z=10$ using a low resolution run, then four 
of the most massive virialized halos are identified using the HOP 
algorithm \citep{hop}. 

Each of the four halos represents a potential site for primordial star
formation. The particles identified with each halo at $z=10$ are 
traced to their positions at $z=99$, and three additional levels of nested 
static subgrids are defined within the root grid such that all halo 
particles are contained within the highest resolution subgrid at all
times (the highest resolution region has an effective resolution
of $1024^3$). The subgrids are centered on a given halo, so that while 
all the simulations share the same root grid and initial conditions, 
the high resolution subgrids are defined independently for each.
The initial conditions are then regenerated to include both baryons 
and dark matter. In the highest resolution subgrid the dark matter 
particle mass is $27.2~\msun$ and the average baryonic 
mass in each cell is $5.6~\msun$.

In the high resolution runs this innermost subgrid is further
refined up to a maximum of 25 levels, with each refinement 
increasing the cell resolution by a factor of two along each axis. 
Cells are flagged for refinement whenever the baryonic or dark 
matter mass in a cell is more than 4 times the resolution
masses quoted above \citep{truelove97}.  In addition, a cell is 
refined if the cell width $\Delta x$ is less than 0.25 times the local 
Jeans length. The dark matter density is smoothed on a scale of 
1 comoving parsec in order to avoid heating of the baryons due 
to discreteness effects of the dark matter particles, as in \citet{abn02}. 
When the simulations reach the maximum refinement level, the cell 
length within the highest refinement region is $46.1$ AU and 
the gas mass resolution is $9.7 \times 10^{-3}~\msun$.

For each halo, two runs are performed, one including HD chemistry 
and one without, to isolate the effects of HD cooling and facilitate 
comparison with previous simulations where HD was not included.
We test the case of primordial star formation in ionized gas as
discussed in the introduction by repeating the simulations for three of 
the halos while ionizing the full simulation volume at $z=20$.

\section{Results}\label{sec:results}

\subsection{Unperturbed Primordial Gas}\label{sec:results1}

\begin{figure*}[!t]
 \epsscale{0.8}
 \plotone{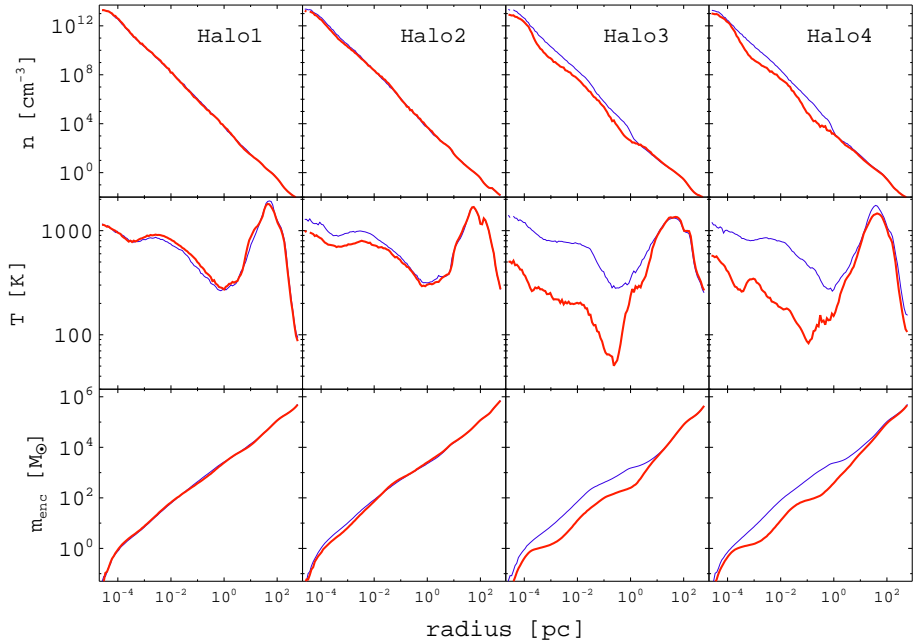}
 \caption{ 
  Radial profiles of gas properties at the final timestep for simulations
  with unperturbed primordial gas.  H$_2$ runs are indicated by thin 
  blue lines and HD runs by thick red lines.  The two more massive 
  halos, Halo1 and Halo2, show little difference between runs with 
  and without deuterium chemistry.  The two less massive halos, 
  Halo3 and Halo4, show significantly lower temperatures within the 
  collapse region, and are also less dense within the inner parsec.
 \label{fig:h2hdcomp1}
 }
\end{figure*}

The first case we consider is for unperturbed primordial gas
(Population III.1), beginning at $z=99$ and evolving normally until 
the formation of a protostar. For all eight runs (4 $\times$ H2 + 4 $\times$ HD) 
a core (proper) density of $n \sim 10^3~\cc$ is reached at  $z\sim15$,
indicating the collapse of a protostellar seed. The H$_2$ mass 
fraction quickly climbs to $10^{-3}$ at a radius of $\sim 1$ pc and 
effectively cools the gas, lowering the temperature to a minimum 
of $\sim 300$K. As the collapse continues, the density exceeds 
$n \sim 10^4~\cc$ and the H$_2$ cooling rate becomes 
independent of density. The temperature of the contracting gas 
rises, and once a density of $10^8~\cc$ is reached H$_2$ is 
formed rapidly through the three-body process until the core is 
fully molecular. The simulations are terminated when the core density 
exceeds $n \sim 10^{13}~\cc$, though we do not consider the inner
core in detail and restrict our discussion to densities $<10^{10}~\cc$
(see \S\ref{sec:chemistry}).
The inner $\sim 20$~AU contains a fully molecular $\sim1 \msun$ 
protostar, consistent with the results of \citet{abn02}.
 
\begin{figure}[htb]
 \epsscale{1.2}
 \plotone{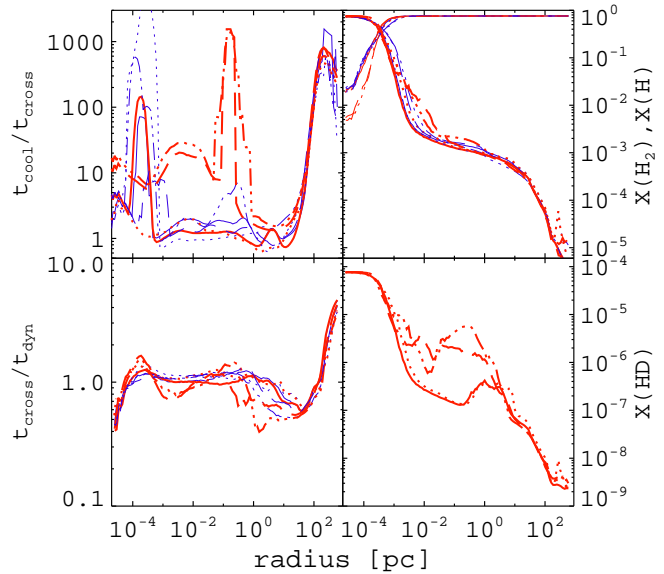}
 \caption{
  Upper left: ratio of cooling time to crossing time for all 8 runs, 
  with H2 runs in thin blue lines and HD runs in thick red lines.  
  Halo1 is shown as solid lines, Halo2 as dotted lines, Halo3 as 
  dashed-dotted lines, and Halo4 as long dashed lines.  The cooling 
  time is much greater in the Halo3-HD and Halo4-HD runs where HD 
  cooling is efficient due to the reduced cooling efficiency at low
  temperatures.
  Lower left: ratio of crossing time to dynamical time, which is very 
  close to one for all cases and indicates dynamical stability.
  Upper right: mass fractions of H$_2$ and H.  The fraction of molecular 
  hydrogen approaches unity within the innermost part of the cloud.
  Lower right: mass fraction of HD, which is enhanced by a factor 
  of 20-50 in the Halo3-HD and Halo4-HD runs.
 \label{fig:h2hdcomp2}
 }
\end{figure}
Properties of the individual halos are given in Table~\ref{tab:haloprop}. 
The four halos span a factor of $\sim 2$ in virial mass. At the virial scale, 
HD has little effect on the halo gas. The collapse to a protostar occurs at 
similar redshifts and virial temperatures whether HD is included or not. 
\begin{figure}[htb]
 \epsscale{1.2}
 \plotone{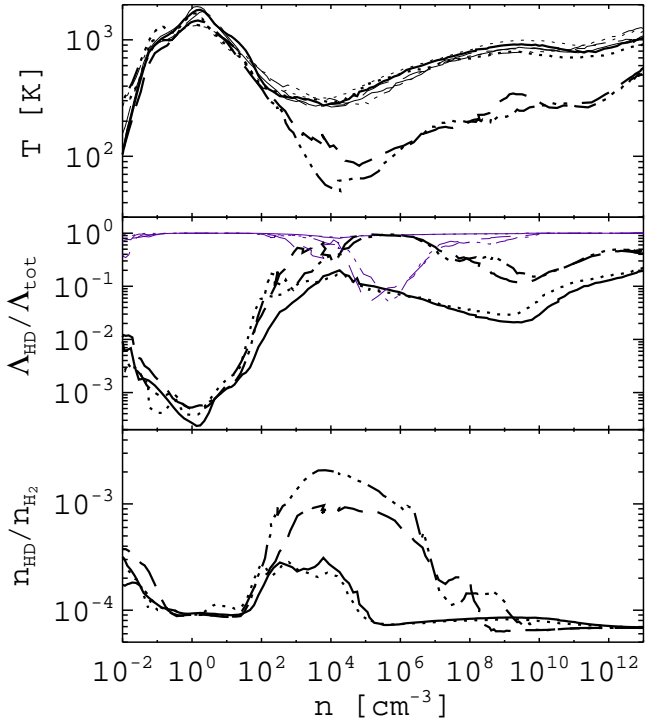}
 \caption{ 
   Contribution of HD to the total cooling in the HD runs. Line styles used 
    for each halo are described Figure~\ref{fig:h2hdcomp2}.
   The top panel shows temperature as a function of density. For reference,
    the results from H2 runs are shown with thin lines.
   The middle panel shows the contribution to the total cooling from HD, with
    thin lines showing the contribution of H$_2$ cooling in the HD runs. 
   The bottom panel shows the HD/H$_2$ ratio, indicating where fractionation 
    has occured.
   For Halo1 and Halo2, HD provides $\la 10\%$ of the total cooling the gas, 
    while for Halo3 and Halo4, the HD/H$_2$ ratio is greater 
    by $3-5\times$ and HD provides $\ga 90\%$ of the total cooling over a wide
    range in density.
  \label{fig:hdcool}
 }
\end{figure}
However, the properties of the halo gas well inside the virial radius, 
particularly in the region where the minimum temperature is reached, 
are clearly affected for two of the four halos, as is evident in the radial 
temperature profiles of all four halos (Figure~\ref{fig:h2hdcomp1}).
Within $\sim3$ parsecs, the H2 and HD runs for Halo3 and Halo4 diverge, 
with the HD runs having lower average densities and temperatures 
compared to the H2 runs.
This indicates that HD cooling is effective in these halos, while for the other 
two halos it is not.

Figure~\ref{fig:h2hdcomp2} shows that throughout most of the 
collapse region, HD production closely follows H$_2$ production 
with a factor of $10^{-4}$ reduction in number density. The exception 
is a region of enhanced HD production in a spherical shell  $\sim1$pc 
from the core, corresponding to the minimum temperature. For Halo1 
and Halo2, the HD/H$_2$ ratio reaches a maximum of 
$\sim 2\times10^{-4}$ in this region, and HD cooling contributes 
less than $20\%$ of the overall cooling, which is governed by H$_2$ 
(Figure~\ref{fig:hdcool}). On the other hand, for Halo3 and Halo4
the maximum HD/H$_2$ ratio is $\sim 10^{-3}$ and HD provides 
more than $90\%$ of the total cooling at densities of 
$10^5 \cc \la n \la 10^7 \cc$. HD cools more effectively than 
H$_2$ at low temperatures; when HD cooling is dominant the 
minimum gas temperature drops from $\sim300$K to $\la100$K. 

Interestingly, in the two HD-cooling halos the overall cooling is reduced, 
as can be seen in the ratio of the cooling time to crossing time
(Figure~\ref{fig:h2hdcomp2}). Near the temperature minimum, the
cooling time rises dramatically for the HD-cooling halos, as the
gas temperature is near the CMB temperature and the cooling rate
goes to zero.  Inside this region, the temperature rises while the HD/H$_2$
ratio drops. Thus in the inner part of the cloud the gas remains cool
($T\ga250$K) but there is less HD, so that the overall cooling remains
low ($t_{\rm cool}/t_{\rm cross} \sim 20$)\footnote{The cooling time is defined
as $E/\dot{E}$ and the crossing time as $r/c_s$.  Both are locally defined within
a grid cell.}. The decrease in gas temperature caused by HD cooling 
actually reduces the total cooling rate, so that the gas will contract 
quasi-statically.

An obvious difference is that the two halos with effective HD cooling 
have the two lowest masses (somewhat below $10^6$ $\msun$),
while the two H$_2$ cooling halos have masses around $10^6$ 
$\msun$ or more.  \citet{ripamonti07} noted that 
halos below $\sim 3\times10^5 \msun$ collapse more slowly 
because H$_2$ is not produced in sufficient quantities to efficiently 
dissipate the heat generated by gravitational contraction. The slower 
collapse allows HD to build up until it becomes a substantial coolant. 
This is similar to the results of \citet{bcl02}, who simulated a
$2\times10^5 \msun$ halo with HD cooling that went through an 
extended phase of quasi-hydrostatic contraction and cooled to
lower temperatures. In both of these studies, when more massive
($>10^6\msun$) halos were tested with and without HD chemistry, 
HD was found to have little effect. We find that halos as massive as 
$8\times10^5 \msun$ generate substantial HD cooling, but do not 
take longer to collapse. In both \citet{ripamonti07} and \citet{bcl02}, 
idealized conditions of a single halo were considered, whereas in our 
simulations the full merger history of a halo is treated self-consistently. 
Thus we must examine the evolution of the gas in more detail to 
understand why HD cooling is effective in low mass halos.

We find that HD cooling in unperturbed primordial gas is a highly sensitive 
process. 
\begin{figure}[!ht]
 \epsscale{1.2}
 \plotone{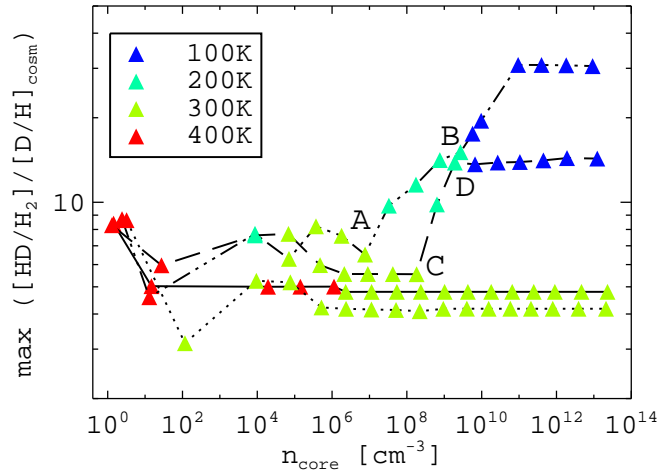}
 \caption{ Evolution of the HD/H$_2$ abundance ratio relative to the 
 cosmological D/H abundance ratio during the collapse (the maximum 
 enhancement within the collapse region is plotted). The x-axis represents 
 the core density, and each halo is followed once a core density of $1~\cc$ is 
 reached. Linestyles are as in Figure~\ref{fig:h2hdcomp2}. Filled symbols 
 correspond to individual output times, with the color representing the 
 minimum temperature in the collapse region at that time. In equilibrium, 
 the HD enhancement would be $\sim 6$ at 250K and $\sim100$ at 100K. 
 The strong HD enhancement in two of the halos only occurs after the 
 core density has exceeded $\sim10^8~\cc$ and the temperature has 
 remained $\la 200$K. For Halo3, the HD enhancement increases from
 6.5 at point A to 15 at point B, a time span of 0.7 Myr.  Halo4 undergoes
 an increase in HD enhancement from 5.5 (point C) to 14 (point D) within 
 0.3 Myr. While Halo3 continues to increase in HD enhancement, for both
 halos this brief period of strong enhancement signals the point at which
 HD cooling goes from marginal to significant importance.
 \label{fig:h2hdevol}
 }
\end{figure}
Two effects combine to drive the effectiveness of HD cooling
in low temperature primordial gas. First, the primary formation channel
for HD,
\begin{equation}
	{\rm D}^{+} + {\rm H}_2 \rightarrow {\rm H}^{+} + {\rm HD},
\end{equation}
is endothermic by 462K, so that HD formation is favored over H$_2$ by a 
factor $e^{462/T}$ \citep{glover07}. Second, the HD cooling
rate per molecule is greater than that of H$_2$ below a critical temperature.
This critical temperature is determined by the gas density and the abundances
of H$_2$ and HD, and is generally in the range 150-250K (see the discussion
and Figure 1 in \citealt{glover07}). 
Should H$_2$ cool the gas to the critical temperature, HD cooling will become 
important and can cool the gas even further. This results in a runaway effect, 
as HD is formed rapidly in the cool gas. 

Indeed, we do not see a slow buildup of HD as in \citet{ripamonti07}; 
rather, the HD enhancement occurs rapidly in the latter stages of the 
collapse. Figure~\ref{fig:h2hdevol} shows the HD enhancement in
each of the four halos, expressed in terms of the HD/H$_2$ abundance 
ratio divided by the cosmological D/H abundance ratio (both values
are ratios of number densities, the cosmological D/H value is stated
in \S\ref{sec:chemistry}). 
\begin{figure}[htb]
 \epsscale{1.2}
 \plotone{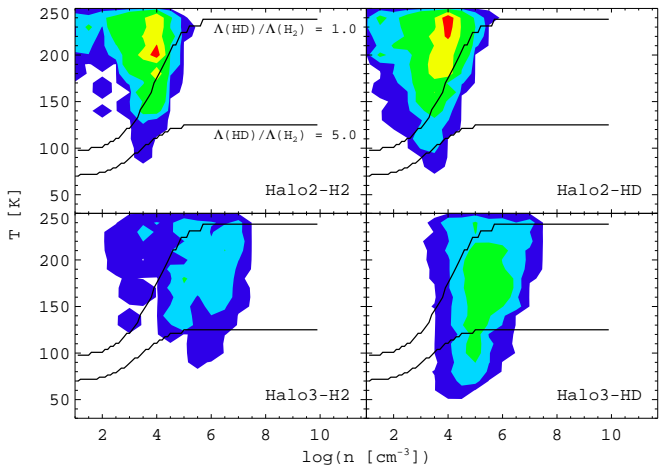}
 \caption{Distribution of the gas in the $n-T$ plane for two halos, 
 at a time when the central density is $\sim5\times10^7~\cc$. 
 Contours represent fractions of $(0.1,2.5,10,100)\times10^{-6}$ 
 of the halo virial gas mass, which is $\sim10^6\msun$ for Halo2 and
 $\sim7\times10^5\msun$ for Halo3. The upper line shows the critical 
 temperature at which the HD cooling rate per unit volume is 
 equal to the H$_2$ cooling rate, calculated at the corresponding 
 density and temperature and assuming that 
 $n_{{\rm H}_2} = 10^{-3}~n_{\rm H}$ and 
 $n_{\rm HD} = 3\times10^{-4}~n_{{\rm H}_2}$, 
 which are typical values for the gas in these regions (see
 Figure~\ref{fig:hdcool}). The lower line shows the the 
 temperature at which HD provides 5 times the cooling of 
 H$_2$. 
 The upper panels show the gas distribution for both the 
 H2 (left panel) and HD (right panel) runs of Halo2.  Most of the gas
 is above the critical region; this halo does not exhibit effective HD 
 cooling.  
 The lower panels show that for Halo3 the cool gas is shifted to higher 
 densities and a much greater amount of gas is within the critical region; 
 this halo does exhibit effective HD cooling.
 \label{fig:contours}
 }
\end{figure}
At an early stage in the collapse, when the core density is $\sim 1\cc$, the 
maximum HD enhancement is $\sim8$ for all halos. When Halo3 reaches 
point A, the minimum temperature is 220K, and the HD enhancement is still 
only 6.5. Within 0.7 Myr, point B is reached, where the minimum temperature 
has dropped to 150K, and the HD enhancement has jumped to 15. Halo4 
undergoes a similar process, at point C the minimum temperature is 225K 
and the HD enhancement is 5.5, but 0.3 Myr later at point D the temperature
has dropped to 100K and the HD enhancement is 14.  Both halos collapse
from $n_{\rm core} = 1\cc$ to $n_{\rm core} = 10^{13}\cc$ in 
$\sim 100$ Myr, thus most of the HD enhancement occurs in less than 
1\% of the total collapse time.  In fact, we find that HD cooling exhibits
a `critical' nature, in the sense that either HD cooling is very important
to the cooling or it is not important at all.  This behavior stems from the
rapid HD fractionation and the runaway effect of cooling.

The conditions for effective HD cooling are then that the gas
must cool by H$_2$ cooling to the critical temperature and remain there
long enough for a buildup of HD to occur. This is most likely to occur in 
lower mass halos. Halos with large gravitational potentials have higher virial
temperatures and more dynamical activity. Mergers from infalling gas 
clumps will disrupt pockets of cool gas. Lower mass halos, which collapse 
more uniformly, are better able to form dense pockets of cool gas where 
HD can form.

We also find evidence for the critical nature of HD cooling by looking in 
detail at the low temperature gas in our simulations. Figure~\ref{fig:contours} 
shows the $n-T$ plane for Halo2 and Halo3 at a time when the central density 
is $5\times10^7~\cc$, just before the large HD enhancement seen in two of 
the halos (Figure~\ref{fig:h2hdevol}). It is clear that the lower mass halo, Halo3, 
has much more gas at higher densities ($\sim10^5\cc$) and low temperatures 
($T<200$K) than Halo2. This gas is well below the critical temperature and is 
cooling effectively by HD. While some gas in the more massive Halo2 is just 
below the critical line, most of it remains above and HD cooling is not effective 
for this halo. Our results, combined with the findings of \citet{ripamonti07} 
and \citet{bcl02}, suggest that a set of conditions must be met for effective HD 
cooling in unperturbed primordial gas, and these conditions are most likely to 
occur in $<10^6\msun$ halos.

\subsection{Primordial Star Formation in Ionized Regions}\label{sec:results2}

The second case we consider is primordial star formation in ionized regions 
(Population III.2). In sites of primordial star formation, molecular 
hydrogen forms primarily through the H$^-$ channel, 
\begin{eqnarray}
	\mathrm{H} + \mathrm{e}^{-} \rightarrow \mathrm{H}^{-} + \gamma \\
	\mathrm{H} + \mathrm{H}^{-} \rightarrow \mathrm{H}_2 + \mathrm{e}^{-}
\end{eqnarray}
a reaction which is catalyzed by free electrons and thus enhanced
in ionized gas. Ionized primordial gas could be found in shocks from
colliding massive halos, or in the photoionized regions surrounding
the first generation of stars. The latter case is of particular interest,
as the end state of $\ga100\msun$ primordial stars is probably either
direct collapse to black holes or complete destruction in pair instability
supernovae \citep{heger03}. 
In either case a sizable HII region will result, 
and as the first stars themselves form in large overdensities these fossil 
HII regions will host dense gas clouds that could be sites for subsequent 
star formation. As H$_2$ forms rapidly in the ionized regions, the gas cools 
to T$\sim100$K, where HD formation is highly favored and HD cooling can 
drive the temperature even lower, essentially to the CMB temperature. The 
resulting low Jeans masses imply lower mass stars \citep{no05,jb06}.

We explore the formation of primordial stars in ionized gas through
the simple approach of ionizing the entire simulation volume at $z=20$.
This technique has been similarly utilized with \enzo\ by \citet{oshea05}
but without HD chemistry. While this method is not equivalent
to a full treatment of radiative transfer, it does allow us to examine in
detail the effect of HD cooling in ionized regions for otherwise identical
halos, while being much less expensive computationally.

\begin{figure}[!bht]
 \epsscale{1.2}
 \plotone{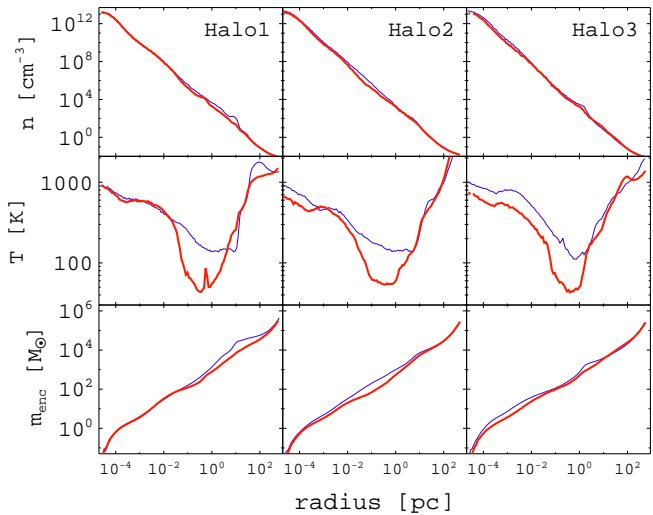}
 \caption{
   The same as Figure~\ref{fig:h2hdcomp1} but for the flash-ionized 
   runs. Similar results are found for the three halos examined, in all 
   cases temperature minimum approaches the CMB temperature at 
   the redshift of collapse.
 \label{fig:flash_h2hdcomp1}
 }
\end{figure}

We used three of the simulation volumes described in \S~\ref{sec:setup}, 
so that the evolution to $z=20$ is identical to the unperturbed case
(\S~\ref{sec:results1}), but the late stage evolution begins with a high
abundance of free electrons. As previously, the simulations are performed
once without HD chemistry and once with HD included.

Table~\ref{tab:flash_haloprop} shows the properties of the halos for the
ionized runs. Compared to the unperturbed case (Table~\ref{tab:haloprop}), 
the collapse for a given halo occurs somewhat earlier and at lower virial 
masses and temperatures in ionized regions.  The resulting density,
temperature and enclosed mass profiles are shown in 
Figure~\ref{fig:flash_h2hdcomp1}. As in Figure~\ref{fig:h2hdcomp1}, the 
HD simulations show distinctly lower temperatures than the H2 runs in the 
range 0.01 - 10 pc, but now the effect is independent of mass.

\begin{figure}[!thb]
 \epsscale{1.2}
 \plotone{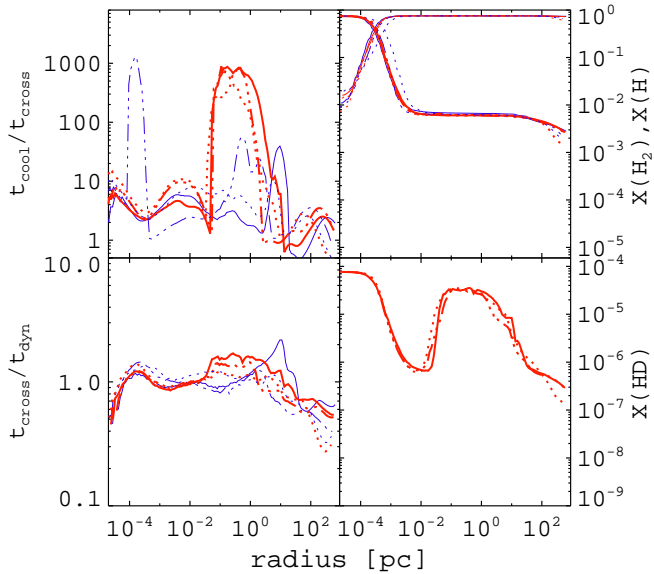}
 \caption{
  The same as Figure~\ref{fig:h2hdcomp2} but for the flash-ionized runs.
  The molecular hydrogen abundance is $> 10^{-3}$ throughout due to the
  enhancement of the H$^-$ channel by the free electrons. The cloud is
  dynamically stable ($t_{\rm cross}/t_{\rm dyn} \sim 1$). HD is greatly
  enhanced within 1pc, where $T\sim100$K.
 \label{fig:flash_h2hdcomp2}
 }
\end{figure}

\begin{figure}[htb]
 \epsscale{1.2}
 \plotone{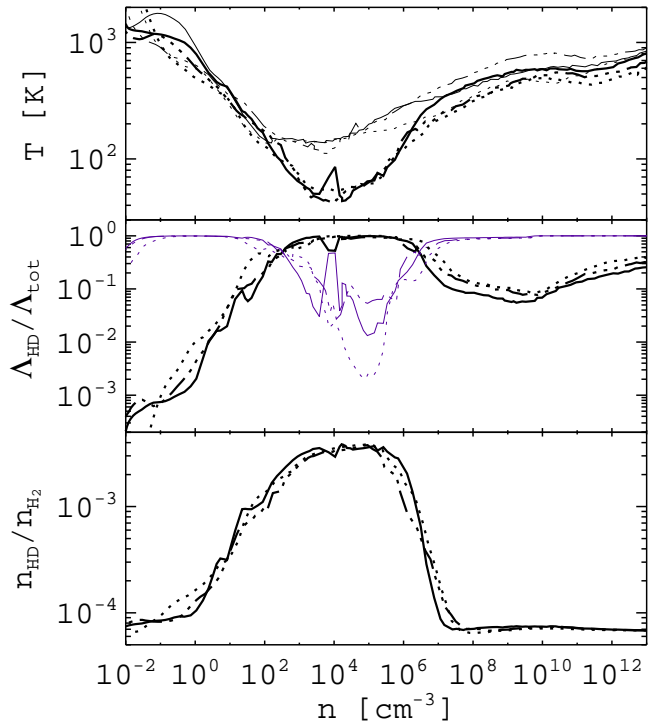}
 \caption{
  The same as Figure~\ref{fig:hdcool} but for the flash-ionized runs.
  Note that HD cooling is now important for all halos, accounting for
  95-99.7\% of the cooling in the $T<100$K region.
 \label{fig:flash_hdcool}
 }
\end{figure}

In Figure~\ref{fig:flash_h2hdcomp2} it is clear that the molecular hydrogen
fraction is greatly enhanced over the unperturbed case as expected. The
extra H$_2$ lowers the gas temperature to $\sim130$K even without 
HD cooling, a factor of $\ga 2$ lower than achieved in the H$_2$ 
cooling unperturbed halos. When HD chemistry is included, it forms rapidly
in the low temperature gas. 
This can be seen in the large bump in the HD fraction at $\sim 1$pc, 
which reaches a maximum of $4\times10^{-5}$, nearly two orders of 
magnitude greater than seen in the unperturbed halos. 
As a result, the temperature is driven to the CMB floor, and HD
provides $>95\%$ of the total cooling in the low temperature zone
(Figure~\ref{fig:flash_hdcool}).

It is interesting to compare the radial development of the H$_2$ and
HD abundances shown in Figure~\ref{fig:flash_h2hdcomp2}.  Moving 
inward from a radius $\sim100$pc, the H$_2$ mass fraction rises to a 
maximum of $\sim6\times10^{-3}$, and remains essentially constant 
until a radius of $\sim10^{-3}$pc. On the other hand, the HD mass fraction 
steadily rises from $\sim100$pc to $\sim1$pc, as the temperature drops from 
$\sim1000$K to $\sim100$K. 
Over the next decade in radius, the temperature 
is roughly constant near the minimum, and the HD mass fraction remains at 
its peak of $\sim3\times10^{-5}$. 
In this region HD fractionation is nearly 
complete, and the HD/H$_2$ ratio reaches $4\times10^{-3}$
(Figure~\ref{fig:flash_hdcool}). At a density of $\sim10^6\cc$, 
the HD cooling function saturates and the temperature rises, 
leading to a sharp drop in the HD abundance, while the H$_2$ 
abundance is still roughly constant. In the inner 0.01pc, H$_2$ 
forms rapidly through the three-body process while the temperature 
again climbs to $\ga700$K, so that the HD abundance again follows 
the H$_2$ abundance reduced by a factor of $7\times10^{-5}$.

\section{Discussion}\label{sec:discussion}

\subsection{General effects of HD cooling}\label{sec:hdeffect}

In general, HD cooling is only important once H$_2$ cooling has
lowered the gas temperature to $\sim200$K. This only occurs
once the gas is already quite dense ($n\sim10^4\cc$). Thus
HD cooling does not affect which halos will form stars, but it does
alter the properties of the gas in the region of the temperature 
minimum.

For halos with effective HD cooling, the minimum temperature is 
not only lowered by a factor of a few compared to runs without HD, 
the location of the minimum is shifted to a smaller radius and hence 
higher densities. When H$_2$ cooling dominates the minimum 
temperature occurs at a density of 
$n \sim 10^3-10^4 \mathrm{cm}^{-3}$. When HD cooling dominates, 
the minimum temperature occurs at a density of 
$n \sim 10^4-10^5 \mathrm{cm}^{-3}$. 

This effect is easily understood by considering the properties of the 
H$_2$ and HD cooling functions. First, it is useful to assume that the 
gas is thermally balanced with heating due to gravitational collapse 
balancing radiative cooling.  The gravitational energy input rate per unit
volume can be approximated as
\begin{equation}
n \left( \frac{GM(r)}{r} \right) t_{\rm dyn}^{-1} \propto n^{3/2}.
\end{equation}
To derive this simplified expression, we have assumed that the 
enclosed mass $M(r)$ is proportional to $r$, or alternately that
$\rho(r) \propto r^{-2}$, as can be seen in Figures~\ref{fig:h2hdcomp1}
and~\ref{fig:flash_h2hdcomp1}. The dynamical time 
$t_{\rm dyn} \propto n^{-1/2}$.

Over a short range in temperature, the cooling function can be roughly 
approximated as
\begin{equation}
  \Lambda \propto \left\{ 
     \begin{array}{ll}
	 n^2 T^\alpha &  (n < n_{\rm cr}) \\
	 n T^\alpha  & (n > n_{\rm cr})
     \end{array}
    \right.
\end{equation}
depending on whether the density $n$ is above or below the critical density 
$n_{\rm cr}$.  For both the H$_2$ and HD cooling curves, $\alpha$ is quite 
large, typically of order 3-5 in the interesting range.  By equating these two 
expressions for the heating and cooling we can derive a relation between 
density and temperature.  When we do this, we see that the temperature 
reaches a minimum at the critical density, where it transitions from 
$T \sim n^{-0.5/\alpha}$ to $T \sim n^{0.5/\alpha}$. The critical density for 
H$_2$ is $n \sim 10^3-10^4 \mathrm{cm}^{-3}$, while for HD it is 
$n \sim 10^5-10^6$ cm$^{-3}$ \citep{lipovka,flower00}. This explains the 
shift in the location of minimum temperature to higher densities as seen 
in Figures~\ref{fig:hdcool}~and~\ref{fig:flash_hdcool} 
(and Tables~\ref{tab:haloprop}~and~\ref{tab:flash_haloprop}). In addition, 
the high value of $\alpha \sim 4$ explains the relatively flat temperature 
profiles seen in the simulations.  The increase in the temperature at densities 
beyond the critical density ultimately leads to a shutoff of HD cooling and, as 
we see in the next section, important implications for the masses of the
resulting stars.

\subsection{Masses of the First Stars}

\begin{figure}[htb]
 \epsscale{1.2}
 \plotone{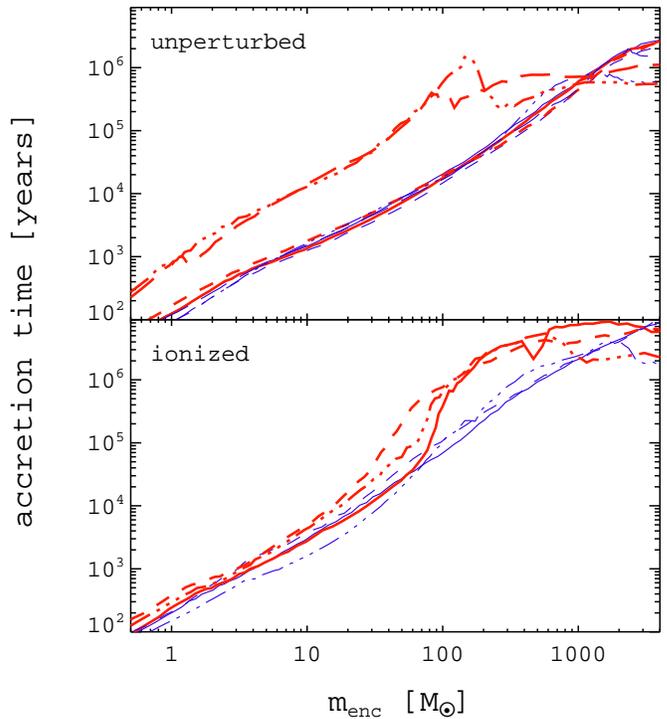}
 \caption{ Accretion times onto protostars in all simulations calculated using
 the Shu rate, with $t_{\rm acc} = m/\dot{m}$. Linestyles are the same as 
 Figure~\ref{fig:h2hdcomp2}. The upper panel shows results from the 
 unperturbed runs, while the lower panel shows results from the ionized runs. 
 In the unperturbed case, HD has little effect on the expected final mass for the 
 stars in Halo1 and Halo2, while having a strong effect for Halo3 and Halo4.
 The accreted mass is reduced by a factor of $\sim6$. In the ionized 
 case, HD cooling reduces the accreted mass by $\la2$. The mass scale
 for primordial halos with effective HD cooling is quite similar to that of
 ionized halos.
 \label{fig:accretion}
 }
\end{figure}

We have shown that HD cooling can be as important as H$_2$ cooling for 
some low mass primordial gas clouds and in ionized regions. We can now 
examine whether HD cooling changes the nature of the collapse enough 
to alter the mass function of the first stars. Figure~\ref{fig:accretion} shows 
the accretion timescales for all simulations, calculated according to the Shu 
isothermal model \citep{Shu1977}, where $\dot{m} = 0.975c_s^3/G$ (we obtain 
similar but noisier results using the instantaneous mass inflow rate, 
$\dot{m}=4\pi r^2v_r \rho$).  

For primordial halos, the accretion times are similar for all of the H$_2$ cooling 
halos, suggesting $\sim 300 \msun$ stars will form after $10^5$~years of 
accretion. However, the two HD cooling halos accrete significantly less mass 
for a given accretion time.  This results from the decreases in temperature and
enclosed mass at $r\sim1$pc when HD cooling is effective. This is an 
important scale for star formation as this is the region at which the 
Bonner-Ebert mass is at a minimum. Our results suggest that stars of 
$\sim50\msun$ will form, much lower than is typically seen in simulations of 
first generation star formation. We conclude, in agreement with \citet{ripamonti07},
that there are two mass scales for the first stars. Some (most) stars will form in 
halos of $10^6\msun$, cool by H$_2$, and have masses $>100\msun$. 
The remainder will form in less massive halos, cool by HD, and have 
masses $<100\msun$. 
According to \citet{heger03}, metal-free stars
with masses $>100\msun$ will produce pair-instability supernovae, while those
with masses $40-100\msun$ will collapse directly to black holes. Thus HD 
cooling may be important in setting the mass scale of the first stars and in the 
later evolution of the interstellar medium.

The inner regions of the ionized halos are quite similar in density and temperature,
but when HD cooling is included the temperature is lower by a factor of $\ga 2$.
The accreted mass is affected by a similar amount; the lower panel of
Figure~\ref{fig:accretion} shows that HD cooling lowers the expected mass of
primordial stars in ionized regions from $\ga100\msun$ to $\sim60\msun$.
This mass scale agrees well with the results of \citet{yoshida07b}, who examined
the development of a protostar in a primordial HII region with SPH simulations
that included HD chemistry. They followed the evolution of the protostar to a central 
density of $10^{18}\cc$ including a detailed protostellar evolution model
that traced the evolution to the zero-age main sequence. They predicted that
a $\sim40\msun$ star would form in the HII region, more massive than earlier 
simulations had predicted. 
Our finding of a similar mass scale for three different halos suggests that while 
second generation primordial stars are not as massive as the first generation, 
they are still quite massive.

We note that extrapolating a final stellar mass from estimations of the
accretion rate is highly uncertain and some caution should be taken
in intrepreting the stellar masses. It is unclear whether a direct relationship
exists between accretion rate and final stellar mass, whether the
accretion is spherical \citep{omukai03} or through an accretion disk
\citep{tan04,McKee2007}.  Taking these uncertainties into consideration,
it is clear that HD cooling has a significant effect on both the temperature
and density evolution of lower mass Population III.1 halos. The total
mass within a given radius is much lower when HD cooling is effective
(Figure~\ref{fig:h2hdcomp1}), implying stars of much lower mass will
form.  In ionized halos, the differences are much smaller
(Figure~\ref{fig:flash_h2hdcomp1}), implying that in this case the resulting 
stars will not be significantly less massive.

We do not see signs of fragmentation in projections of the inner collapse regions.
We also do not see evidence for the formation of rotationally-supported disks,
as $v_{\rm circ}/v_{\rm kep} < 1$ in all our simulations. 
This is not a surprise, as we do not expect HD cooling to cause fragmentation 
-- like H$_2$ it has a steep cooling function and so is stable to thermal instabilities.  
In addition, as we see in Figure~\ref{fig:flash_h2hdcomp2}, the cooling time never 
drops below the dynamical time, another important criterion for fragmentation.

\subsection{Rate uncertainties}

The results presented here depend on the accuracy of the chemical model.  We
showed in \S~\ref{sec:chemistry} that the chemistry solver numerically reproduces
the expected abundances in equilibrium; however, the calculated values are only
as accurate as the rates given as input.  \citet{Glover2006} and \citet{glover07} 
have noted that two of the rates most strongly tied to H$_2$ formation are also the 
most uncertain \citep[see also][]{glover08}. The first, the H$^-$ channel, is less relevant 
to the unperturbed case but may be important in ionized regions.  This uncertainty will 
propagate into simulations with HD, as HD primarily forms through H$_2$.  As discussed 
in \citet{Glover2006}, this can be an important source of uncertainty in these calculations, 
and we reiterate the need for improved rate measurements and calculations.

The second uncertainty discussed by \citet{glover07} is in the three-body H$_2$ 
formation process.  This uncertainty will not qualitatively affect our results, as the
three-body process is only important in the inner core of the collapse region, 
where the temperature climbs up to $\ga500$K and HD is less important.  We 
did not include an analogous three-body process for HD formation, though for 
the same reason we do not expect this to be an important process.  

Another source of uncertainty is in the cosmological D/H ratio.  
As discussed in \S~\ref{sec:chemistry}, there is a factor $\sim2$ uncertainty
in cosmological measurements of this value.  Figure~\ref{fig:h2hdevol} shows
that HD cooling halos undergo a rapid fractionation phase; this is unlikely to be
sensitive to small changes in the cosmological deuterium abundance.  In addition,
the final amount of HD enhancement differs by a factor of 2 for the HD cooling
halos, yet the fraction of cooling from HD is similar for both cases 
(Figure~\ref{fig:h2hdevol}).


\section{Conclusions}\label{sec:conclusions}

In this paper, we have implemented a minimal chemical network for tracking
the formation of HD in the \enzo\ cosmological hydrodynamics code.  We have
verified the network using both equilibrium and non-equilibrium tests and then
applied this to the formation of primordial stars in high-redshift halos with
cosmological initial conditions.  We examined a range of halos under
two different assumptions for the initial electron fraction.  In the first case,
the halos were unperturbed, with an initial electron fraction as predicted by
standard cosmological evolution in the absence of ionizing backgrounds.
As in previous work, the first collapsed objects form in $10^{5-6}$ $\msun$ halos
at $z \sim 15$.  In the second case, we ionized the halos at $z=20$ to investigate
the impact of free electrons on the collapse.  Although this simple `flash'
ionization is not a realistic description of, say, a nearby photoionizing source,
it does allow us to investigate the impact of HD cooling in the presence
of a source of free electrons.
Our primary results are as follows.

\begin{itemize}

\item For Population III.1 star formation in high mass halos (roughly $10^6$ $\msun$ 
and above), HD cooling is unimportant.  However, for the lower mass halos in our 
sample there was a factor of 10-100 increase in the HD fraction (compared to the high 
mass halos).  This, is turn, caused HD cooling to dominate H$_2$ cooling over the 
density range $10^4$ cm$^{-3}$ to $10^7$ cm$^{-3}$ (corresponding approximately 
to $10^{-2}$ pc to 1 pc, in radii).  The HD cooling permitted gas to reach nearly down
to the CMB temperature.  

\item We found that in order for HD cooling to be important, a sufficient amount of gas 
needed to be at a low enough temperature and a sufficiently high density for HD to form 
preferentially over H$_2$. However, once this critical line was crossed, rapid HD 
fractionation coupled with runaway cooling quickly led HD to dominate the cooling rate.  
This requirement for lower temperatures helps to explain why low mass halos formed 
HD preferentially over high mass halos.

\item For the ionized case (Population III.2), HD cooling was important in all halos
examined regardless of mass.   HD cooling was efficient enough for gas to reach 
the CMB temperature over an even larger range of radii.

\item In all cases, once the density increased above the HD critical density (around 
$10^{5-6}$ cm$^{-3}$), the temperature began to rise again and invariably increased 
by about $\sim 500$ K, leading to the suppression of HD relative to H$_2$.  This means 
that the interior solution ($r < 10^{-2}$ pc) relies only on H$_2$ cooling and so looks
quite similar to the no-HD case.  In addition, no sign of fragmentation was found.

\item Despite the fact that HD cooling led to lower temperatures (often down to the
 CMB temperature), we found that the impact on the predicted mass accretion rates 
 was relatively mild.  It was largest for the unperturbed case, where the low mass halos 
 were predicted to form stars a factor of six times lower than the no-HD case (from 
 $\sim 300\msun$ to $\sim 50$ $\msun$, assuming an accretion time of $10^5$ years
 -- see \S\ref{sec:discussion} for a discussion of the uncertanties in accretion mass
 estimates). For the ionized case, the predicted accretion rates were nearly identical 
 with and without HD, leading to at most a factor of two reduction in the mass of 
 the predicted stars forming out of ionized regions (from $\sim 100$ $\msun$ to 
 $\sim 50$ $\msun$).

\end{itemize}

These results have a number of interesting implications.  First, they suggest a way 
for at least some relatively lower mass stars to form in primordial (or near primordial) 
halos, as some interpretations of the observed extremely low metallicity halo stars 
have required \citep{Tumlinson2007, Komiya2007}.  However, they also indicate that 
HD cooling by itself will not form substantially lower mass stars (e.g. by a factor of ten) 
as some previous analytic work has suggested.  In particular, we have not found any 
sign of a new class of Population III.2 objects (at least for the initial conditions surveyed).

Our work suggest two avenues of future research.  The first is a more careful and 
systematic investigation of the conditions for HD cooling in ``unperturbed'' halos.  
While our results indicate that HD cooling is more important in low mass halos, it is 
not clear how this depends on merger histories or stochastic effects.  More work is 
required to find the fraction of halos which form via the HD channel for a complete 
sample of halos.  In the ionized case, our methodology for introducing free electrons 
was deliberately simple.  It would be useful to examine more realistic ionization 
modalities \citep[e.g.,][]{Wise2007,yoshida07a}.

\acknowledgements

GB acknowledges support from NSF grants AST-05-07161, AST-05-47823, and
AST-06-06959, as well as computational resources from the National
Center for Supercomputing Applications.  We would like to thank the referee
for many helpful suggestions which greatly improved the clarity of the paper.
We also thank Tom Abel, Simon Glover, Zoltan Haiman, Mordecai Mac-Low, 
and Dan Whalen for useful discussions.

\clearpage
\appendix



In this appendix we show the derivation of the analytical formulas used to test 
the equilibrium abundances of the chemical network, as described in
\S~\ref{sec:chemistry} and shown in Figure~\ref{fig:equilibrium}. The 
chemical network for all reactions not involving deuterium is taken from the 
work of \citet{abel97}.  The complete set of reactions and their associated 
rates are given in that work, for brevity we focus here on the species most 
relevant to cooling in the primordial gas. For ease of comparison, rates 
listed in this appendix are also denoted R\#, with the number corresponding 
to the rate number used in \citet{abel97} (and also within the \enzo\ source code).

The equilibrium abundance of ionized hydrogen is calculated from the collisional
balance of H$^+$ (\citet{abel97} rates R1-R2):



\begin{eqnarray}
	{\rm H} + e^- & \rightarrow & {\rm H}^+ + 2e^- ,  \\
	{\rm H}^+ + e^- & \rightarrow & {\rm H} + h\nu .
\end{eqnarray}

By assuming $n_e = n_{\mathrm{H}^+}$, we derive
\begin{eqnarray}
	\nHp = \frac{k_1}{k_2}\nH .
\end{eqnarray}

The abundance of neutral hydrogen is found by assuming 
$n_{\rm H,tot} = \nH + \nHp$.

The two processes for creating molecular hydrogen include the dominant
${\rm H}^-$ channel and charge exchange with ${\rm H}_2^+$ (rates R8 and R10),
\begin{eqnarray}
	{\rm H} + {\rm H}^- & \rightarrow & {\rm H}_2 + e^- , \\
\label{rate:r10}
	{\rm H}_2^+ + {\rm H} & \rightarrow & {\rm H}_2 + {\rm H}^+ .
\end{eqnarray}
The three processes governing the destruction of H$_2$ are the reverse of
the above charge exchange reaction, as well as collisional dissociations by
electrons and neutral hydrogen (R11-R13),
\begin{eqnarray}
\label{rate:r11}
	{\rm H}_2 + {\rm H}^+ & \rightarrow & {\rm H}_2^+ + {\rm H} , \\
	{\rm H}_2 + e^- & \rightarrow & 2{\rm H} + e^- , \\
\label{rate:r13}
	{\rm H}_2 + {\rm H} & \rightarrow & 3{\rm H} .
\end{eqnarray}

Thus the equilibrium abundance of H$_2$ is given by
\begin{equation}
	 \nHH = \frac{k_8\nH\nHm + k_{10}\nH\nHHp}
                        {k_{11}\nHp + k_{12}n_e + k_{13}\nH}.
	\label{eqn:h2eq}
\end{equation}

Although the H$_2$ abundance in the low-temperature gas ($T < 10^4$K) typical
of primordial star formation is dominated by the H$^-$ channel, we do not neglect
H$_2^+$ in our equilibrium calculation as it is important at higher temperatures.
In addition to reaction (\ref{rate:r11}) (also denoted R11), the following two reactions produce H$_2^+$ (R9 and R17):

\begin{eqnarray}
	{\rm H} + {\rm H}^+ \rightarrow {\rm H}_2^+ + h\nu , \\
\label{rate:r17}
	{\rm H}^- + {\rm H}^+ \rightarrow {\rm H}_2^+ + e^- .
\end{eqnarray}
H$_2^+$ is destroyed by reaction (\ref{rate:r10}) (also denoted R10) and the following reactions (R18, R19):
\begin{eqnarray}
	{\rm H}_2^+ + e^- & \rightarrow &  2{\rm H} , \\
	{\rm H}_2^+ + {\rm H}^- &  \rightarrow & {\rm H}_2 + {\rm H} .
\end{eqnarray}

The equilibrium abundances for H$_2$ and H$_2^+$ are found by solving
equation~(\ref{eqn:h2eq}) simultaneously with
\begin{equation}
	 \nHHp = \frac{k_9\nH\nHp + k_{11}\nHp\nHH + k_{17}\nHm\nHp}
                        {k_{10}\nH + k_{18}n_e + k_{19}\nHm}.
	\label{eqn:h2peq}
\end{equation}

H$^-$ is an important intermediary in the formation of H$_2$. It is formed
primarily by photo-attachment of H and $e^-$ (R7),
\begin{eqnarray}
	{\rm H} + e^- &  \rightarrow & {\rm H}^- + h\nu .
\end{eqnarray}
H$^-$ is destroyed in a series of collisional processes, including 
reaction (\ref{rate:r17}) (also denoted R17)
and the following rates (R14-R16):
\begin{eqnarray}
	{\rm H}^- + {\rm H} & \rightarrow &  {\rm H} + 2e^- , \\
	{\rm H}^- + {\rm H} &  \rightarrow & 2{\rm H} + e^- , \\
	{\rm H}^- + {\rm H}^+ &  \rightarrow & 2{\rm H} .
\end{eqnarray}
This leads to the equilibrium equation
\begin{equation}
	\nHm = \frac{k_7\nH n_e }
	       { k_8\nH + k_{14}n_e + k_{15}\nH + k_{16}\nHp + k_{17}\nHp } .
	\label{eqn:Hmeq}
\end{equation}

From Table~\ref{tab:rates}, D$^+$ is primarily created by collisional ionization 
and destroyed by recombination and collision with neutral hydrogen,
\begin{equation}
	 \nDp = \frac{k_{50}\nD\nHp}{k_2\nHp + k_{51}\nH}.
	\label{eqn:Dpeq}
\end{equation}

HD is created and destroyed in four exchange reactions involving H$_2$
(Table~\ref{tab:rates}),
\begin{equation}
	 \nHD = \frac{k_{52}\nDp\nHH + k_{54}\nD\nHH}
	                   {k_{53}\nHp + k_{55}\nH}.
	\label{eqn:HDeq}
\end{equation}

We note that the rate for collisional destruction of H$_2$ by H (rate (\ref{rate:r13}), or R13)
used in \enzo\ is significantly higher than the rate for the same process used by
\citet{bcl02} for the temperature range $10^4 \la T \la 10^7$. The equilibrium
abundance of H$_2$ at $T \sim 15000$ K in our calculation is lower by a factor
of 6 than that of \citet[see their Figure 1]{jb06}. Simulations of primordial star
formation do not reach such high temperatures so this difference is unimportant
to the work presented here.

\clearpage

\begin{deluxetable}{rllr} 
\footnotesize
\tablewidth{0pt}
\tablecaption{Reaction rates for deuterium species \label{tab:rates}}
\tablecolumns{4}
\tablehead{
\colhead{} &
\colhead{} &
\colhead{Rate Coefficient} &  
\colhead{} \\ 
\colhead{} &
\colhead{Reaction} &  
\colhead{(cm$^{3}$~s$^{-1}$)} &
\colhead{Reference}
 } 
\startdata
D1 & D + $e^{-}$ $\rightarrow$ D$^{+}$ + 2$e^{-}$ & 
        See expression in reference & 
        A97/1\tablenotemark{a}
        \nl
D2 & D$^{+}$ + $e^{-}$ $\rightarrow$ D + $\gamma$ &
        See expression in reference & 
        A97/2\tablenotemark{a}
        \nl
D50 & D + H$^{+}$ $\rightarrow$ D$^{+}$ + H  & 
        $2.00\times10^{-10}T^{0.402}\mathrm{e}^{-37.1/T}
          - 3.31\times10^{-17}T^{1.48}$ & 
        GP02/5
         \nl
D51 & D$^{+}$ + H $\rightarrow$ D + H$^{+}$ & 
        $2.06\times10^{-10}T^{0.396}\mathrm{e}^{-33.0/T}
          + 2.03\times10^{-9}T^{-0.332}$ & 
        GP02/6\tablenotemark{b}
         \nl
D52 & D$^{+}$ + H$_{2}$ $\rightarrow$ H$^{+}$ + HD & 
        $10^{-9}\times[0.417+0.846 \log{T} -0.137(\log{T})^2]$ & 
        GP02/2
        \nl
D53 & HD + H$^{+}$ $\rightarrow$ H$_{2}$ + D$^{+}$ & 
        $1.1\times10^{-9}\mathrm{e}^{-488/T}$ & 
        GP02/4
         \nl
D54 & D + H$_{2}$ $\rightarrow$ H + HD & 
        $1.69\times10^{-10}\mathrm{e}^{(-4680/T+198800/T^2)}$ & 
        GP02/1\tablenotemark{c}
        \nl
D55 & HD + H $\rightarrow$ H$_{2}$ + D & 
        $5.25\times10^{-11}\mathrm{e}^{(-4430/T + 173900/T^2)}$ & 
        GP02/3\tablenotemark{c}
        \nl
\enddata
 \tablecomments{
The references are \citealt{abel97} (A97) and \citealt{gp02} (GP02), with the
rate numbers used in those papers shown after the reference.  The
first column gives the rate numbers according to the system used internally 
within the \enzo\ code.
  }
  \tablenotetext{a}{The collisional ionization and recombination rates for 
 deuterium are taken from the equivalent rates for hydrogen.}
  \tablenotetext{b}{Rate 6 from \citealt{gp02} contains a typo and was 
 obtained from the original reference, \citealt{savin}.}
  \tablenotetext{c}{\citet{ripamonti07} noted that these rates 
 become unphysically large at $T \le 100$K, following the example of that 
 paper we set the rates at $T \le 100$K to be the same as the rate at
 $T=100$K.}
\end{deluxetable}

\begin{deluxetable}{lrrrrrr}
 \tablecaption{Properties of the simulation halos in the unperturbed runs\label{tab:haloprop}}
 \tablewidth{0pt}
 \tablehead{ \colhead{Run} & 
                    \colhead{$z_{coll}$} &
                    \colhead{$M_{vir}$} & 
                    \colhead{$R_{vir}$} & 
                    \colhead{$T_{vir}$} &
                    \colhead{$T_{min}$} &
                    \colhead{$n(T_{min})$} \\
                    & 
                    &
                    \colhead{($10^5 \msun$)} & 
                    \colhead{(pc)} & 
                    \colhead{(K)} &
                    \colhead{(K)} &
                    \colhead{($10^3 \mathrm{cm}^{-3}$)}
                  }
 \startdata
Halo1-H2 &  14.55 &  10.84 & 213.0 & 1610 &  268 &    5.5 \\
Halo1-HD &  14.71 &  10.24 & 207.0 & 1566 &  286 &    4.1 \\
Halo2-H2 &  15.35 &   9.62 & 194.7 & 1563 &  322 &    3.6 \\
Halo2-HD &  15.34 &   9.82 & 196.2 & 1585 &  300 &    3.2 \\
Halo3-H2 &  14.88 &   6.72 & 177.9 & 1195 &  290 &   31.6 \\
Halo3-HD &  14.74 &   6.94 & 181.5 & 1211 &   66 &   23.7 \\
Halo4-H2 &  14.56 &   9.37 & 202.9 & 1462 &  268 &    5.0 \\
Halo4-HD &  15.02 &   7.93 & 186.4 & 1347 &  103 &   85.2 \\
 \enddata
 \tablecomments{
  The runs are labeled ``H2'' to indicate a run without deuterium chemistry
  and ``HD'' when deuterium is included; they are otherwise identical.
  The redshift of collapse is determined as the redshift when the central
  density exceeds $10^{10} \cc$; the virial mass, radius, and temperature
  are calculated at this time as well.  The minimum temperature $T_{min}$
  is calculated at the final timestep and found by fitting a cubic polynomial
  to the $n$ - $T$ plane in logarithmic space, in the region around the
  temperature minimum (e.g., the upper panel of Figure~\ref{fig:hdcool}).  
  The value of $T_{min}$ and corresponding density $n(T_{min})$ are 
  determined from the minimum of the fitting function, and illustrate the effect 
  of HD cooling in the region of lowest temperature.
}
\end{deluxetable}

\begin{deluxetable}{lrrrrrr}
 \tablecaption{Properties of the simulation halos in the ionized runs\label{tab:flash_haloprop}}
 \tablewidth{0pt}
 \tablehead{ \colhead{Run} & 
                    \colhead{$z_{coll}$} &
                    \colhead{$M_{vir}$} & 
                    \colhead{$R_{vir}$} & 
                    \colhead{$T_{vir}$} &
                    \colhead{$T_{min}$} &
                    \colhead{$n(T_{min})$} \\
                    & 
                    &
                    \colhead{($10^5 \msun$)} & 
                    \colhead{(pc)} & 
                    \colhead{(K)} &
                    \colhead{(K)} &
                    \colhead{($10^3 \mathrm{cm}^{-3}$)}
                  }
 \startdata
Halo1-H2 &  13.53 &  12.72 & 240.5 & 1673 &  132 &    1.3 \\
Halo1-HD &  15.06 &   7.38 & 181.5 & 1286 &   55 &    9.1 \\
Halo2-H2 &  18.22 &   2.21 & 101.5 &  690 &  132 &    2.2 \\
Halo2-HD &  18.74 &   1.93 &  94.4 &  647 &   61 &   12.5 \\
Halo3-H2 &  17.05 &   2.29 & 109.3 &  662 &  132 &    2.1 \\
Halo3-HD &  15.37 &   4.57 & 151.8 &  953 &   58 &    9.4 \\
 \enddata
 \tablecomments{See Table~\ref{tab:haloprop} for a description of the
 table format.}
\end{deluxetable}

\end{document}